\documentclass[aps,pre,showpacs,floats,twocolumn]{revtex4}
\usepackage{amssymb}
\usepackage{graphics}
\usepackage[dvipdfm]{graphicx}
\usepackage{epsfig}
\usepackage{dcolumn}
\usepackage{amsmath}

\begin{document}

\title{Statistical Mesoscopic Hydro-Thermodynamics: The Description of
Kinetics and Hydrodynamics of Nonequilibrium Processes in Single
Liquids}
\author{Jos\'{e} G. Ramos$^{1}$, Cl\'{o}ves G. Rodrigues$^{2}$\footnote{corresponding author:
cloves@pucgoias.edu.br}, Carlos A. B. Silva$^{3}$, Roberto
Luzzi$^{1}$} \affiliation{$^{1}$Condensed Matter Physics Department,
Institute of Physics ``Gleb Wataghin'' State University of
Campinas-Unicamp, 13083-859 Campinas, SP,
Brazil\\
$^{2}$Materials Science Group, Pontifical Catholic University of
Goi\'{a}s,
74605-010 Goi\^{a}nia, Goi\'{a}s, Brazil\\
$^{3}$Departamento de F\'{\i}sica, Instituto Tecnol\'{o}gico de
Aeron\'{a}utica, 12228-901, S\~{a}o Jos\'{e} dos Campos, SP, Brazil}
\date{\today }

\begin{abstract}
Hydrodynamics, a term apparently introduced by Daniel Bernoulli
(1700-1783) to comprise hydrostatic and hydraulics, has a long
history with several theoretical approaches. Here, after a
descriptive introduction, we present so-called mesoscopic
hydro-thermodynamics, which is also referred to as higher-order
generalized hydrodynamics, built within the framework of a
mechanical-statistical formalism. It consists of a description of
the material and heat motion of fluids in terms of the corresponding
densities and their associated fluxes of all orders. In this way,
movements are characterized in terms of intermediate to short
wavelengths and intermediate to high frequencies. The fluxes have
associated Maxwell-like times, which play an important role in
determining the appropriate contraction of the description (of the
enormous set of fluxes of all orders) necessary to address the
characterization of the motion in each experimental setup. This
study is an extension of a preliminary article: Physical Review E
\textbf{91}, 063011 (2015).
\end{abstract}

 \maketitle

%%%%%%%%%%%%%%%%%%%%%%%%%%%%%%%%%%%%%%%%%%%%%%%%%%%%%%%%%%%%%%%%%%%%%%%%%%
%%%%%%%%%%%%%%%%%%%%%%%%%%%%%%%%%%%%%%%%%%%%%%%%%%%%%%%%%%%%%%%%%%%%%%%%%%
%%%%%%%%%%%%%%%%%Section 1%%%%%%%%%%%%%%%%%%%%%%%%%%%%%%%%%%%%%%%%%%%%%%%%
%%%%%%%%%%%%%%%%%%%%%%%%%%%%%%%%%%%%%%%%%%%%%%%%%%%%%%%%%%%%%%%%%%%%%%%%%%

\section{Introduction}

The name \emph{hydrodynamics} was apparently first introduced by
Daniel Bernoulli (1700-1783) to comprise the disciplines of
hydrostatics and hydraulics. Leonard Euler (1707-1783) developed the
equations of motion of a perfect fluid and the accompanying
mathematical theory. Stokes (1819-1903) derived the equations of
motion of a viscous fluid and can be considered to be a founder of
the modern theory of hydrodynamics [1].

Microscopic descriptions of hydrodynamics, that is, the derivation
of kinetic equations from classical or quantum mechanics with the
kinetic (or transport) coefficients written in terms of correlation
functions, is a traditional long-standing problem. An important
aspect is the derivation of constitutive laws that express
thermodynamic fluxes (or currents in the case of matter and energy)
in terms of appropriate thermodynamic forces (typically gradients of
densities in the case of matter and energy). In their most general
form, these laws are nonlocal in space and noninstantaneous in time.
The nonlocality is usually addressed by spatial Fourier transforms,
and then the laws -- expressed in reciprocal space - become
dependent on the wavevector $\mathbf{Q}$. The well-known expressions
of classical (or Onsagerian) hydrothermodynamics are obtained by
performing the limit of $\mathbf{Q}$ going to zero (long
wavelengths). The expressions are then valid in this limit and to go
beyond, it is necessary to introduce a proper dependence on
$\mathbf{Q}$ that is valid, in principle, for intermediate and short
wavelengths (intermediate to large wavenumbers). In phenomenological
theories, this corresponds to going from classical (Onsagerian)
irreversible thermodynamics to extended irreversible thermodynamics
[2,3]. This is what has been called \emph{generalized
hydrodynamics}, that is, to go beyond traditional hydrodynamics, the
latter is restricted to fluctuations occurring at long wavelengths
and low frequencies. This idea has been extensively debated for
decades by the statistical mechanics community. Several approaches
have been used, and a description can be consulted in chapter 6 of
the classic book on the subject by Boon and Yip [4]. Here, we
present a statistical approach based on a grand canonical ensemble
generalized to cover the case of fluids arbitrarily away from
equilibrium and at intermediate to short wavelengths.

Nonlocal effects for describing motion with the influence of
decreasing wavelengths (going towards the very short limit) have
been introduced in terms of expansions in increasing powers of the
wavenumber, which currently consists of what is sometimes referred
to as higher-order hydrodynamics (HOH) or mesoscopic
hydro-thermodynamics (MHT) [5].

Earlier attempts to perform such expansions are the so-called
Burnett and super-Burnett approaches, for the case of mass motion,
and the Guyer-Krumhansl approach [6] in the case of propagation of
energy in semiconductors. The usual approach is based on the moment
solution procedure of the Boltzmann equation, as in the work of Hess
[7], using a higher-order Chapman-Enskog solution method. The
Chapman-Enskog method provides a solution to the Boltzmann equation
consisting of a series in powers of the Knudsen number, $K_{n}$,
given by the ratio between the mean free path of the particles and
the scale of the variation (relevant wavelengths in the motion) of
the hydrodynamic fields. Retaining the linear term in $K_{n}$ gives
the Navier-Stokes equation, the term containing $K^{2}_{n}$
introduces the so-called Burnett correction, and the higher order
terms ($K^{3}_{n}$ and up) give the super-Burnett corrections [8].

The satisfactory development of MHT is highly desirable for covering
a large class of hydrodynamic situations and, in the last instance,
for obtaining insights into technological and industrial processes
with an associated economic interest: for example, oil recovery,
pollution decontamination or CO$_{2}$ sequestration in soils that
are important environmental processes. Indeed, the nonlocal terms
become particularly important in miniaturized devices with
constrained geometries, for example, the nanometer scales in
electronic and optoelectronic devices.

We have considered the question of going beyond earlier approaches
by means of a nonequilibrium statistical-mechanical formalism
[9-22], which thus contains the quantum -- or classical -
microscopic dynamics and the macroscopic nonequilibrium
thermodynamics [23-26], with the equations of hydrodynamics
following from the mechanical equations of motion averaged over a
nonequilibrium statistical ensemble, which is provided by the
associated kinetic theory of the formalism [12-15,27-34].

MHT was derived based on the method of moments for the solution of
the single-particle (or quasi-particle) kinetic equation [35,36]
(generically referred to as the Boltzmann equation), initiated by
Grad [37]. The case of a dilute solution of many Brownian particles
was reported in Ref. [38]. MHT of quasi-particle phonons was
presented under different conditions in Refs. [39-42], and the case
with complex structure fluids (with ``hidden constraints") was
presented in Ref. [43].

In this paper, we present an extensive description of the MHT of a
classical fluid of molecules with internal interactions.
Generalizations of the so-called Maxwell time arise in the
treatment, the issue of the contraction of the description (of
general MHT involving many fluxes of all orders) is discussed, and
the theory is illustrated with an MHT description of order 2.

%%%%%%%%%%%%%%%%%%%%%%%%%%%%%%%%%%%%%%%%%%%%%%%%%%%%%%%%%%%%%%%%%%%%%%%%%%
%%%%%%%%%%%%%%%%%%%%%%%%%%%%%%%%%%%%%%%%%%%%%%%%%%%%%%%%%%%%%%%%%%%%%%%%%%
%%%%%%%%%%%%%%%%%Section 2%%%%%%%%%%%%%%%%%%%%%%%%%%%%%%%%%%%%%%%%%%%%%%%%
%%%%%%%%%%%%%%%%%%%%%%%%%%%%%%%%%%%%%%%%%%%%%%%%%%%%%%%%%%%%%%%%%%%%%%%%%%

\section{Theoretical Background}

Let us consider a fluid of $N$ molecules (with density $n$) of mass
$m$, with $\mathbf{r}_{j}$ and $\mathbf{p}_{j}$ being the coordinate
and linear momentum of the $j$-th molecule in contact with an
external reservoir at temperature $T_{0}$. We write the Hamiltonian
as
%
%Eq. (1)
\begin{equation}
\hat{H} = \hat{H}_{0} + \hat{H}^{\prime} \, ,  \label{Eq. 1}
\end{equation}
where
%
%Eq. (2)
\begin{equation}
\hat{H}_{0} = \sum_{j=1}^{N} \frac{p_{j}^{2}}{2m} \, ,  \label{Eq.
2}
\end{equation}
is the kinetic energy, and
%
%Eq. (3)
\begin{equation}
\hat{H}^{\prime} = \frac{1}{2} \sum_{j \neq k} V(\mathbf{r}_{j} -
\mathbf{r}_{k}) \, ,  \label{Eq. 3}
\end{equation}
represents the pair interaction of the molecules.

A macroscopic (thermodynamic) description is made in terms of the
non-equilibrium statistical ensemble formalism (NESEF) described in
Refs. [9-22], the associated NESEF kinetic theory in Refs. [27-34],
and the NESEF-nonequilibrium thermodynamic theory in Refs. [23-26].
A very briefly summary is given in Appendix A.

The nonequilibrium statistical operator, $\varrho_{\varepsilon}(t)$
in Appendix A, is dependent on a set of microdynamical variables,
which for the present problem are the hydrodynamical variables
consisting of the density of particles
%
%Eq. (4)
\begin{equation}
\widehat{n}(\mathbf{r}) = \int
d^{3}p\,\widehat{n}_{1}(\mathbf{r},\mathbf{p}) \, ,  \label{Eq. 4}
\end{equation}
and the density of energy
%
%Eq. (5)
\begin{equation}
\widehat{h}(\mathbf{r}) = \int d^{3}p \frac{p^{2}}{2m}
\widehat{n}_{1}(\mathbf{r},\mathbf{p}) \, ,  \label{Eq. 5}
\end{equation}
where
%
%Eq. (6)
\begin{equation}
\widehat{n}_{1}(\mathbf{r},\mathbf{p}) = \sum_{j=1}^{N} \delta
\left( \mathbf{r} - \mathbf{r}_{j} \right) \delta \left( \mathbf{p}
- \mathbf{p}_{j} \right) \label{Eq. 6}
\end{equation}
is the single-particle dynamical operator, and the fluxes of all
orders of these densities are,
%
%Eq. (7)
\begin{equation}
\widehat{\mathbf{I}}_{n}(\mathbf{r}) = \int
d^{3}p\,\frac{\mathbf{p}}{m} \widehat{n}_{1}(\mathbf{r},\mathbf{p})
\, , \label{Eq. 7}
\end{equation}
%
%Eq. (8)
\begin{equation}
\widehat{I}_{n}^{[\ell ]}(\mathbf{r}) = \int d^{3}p \,
\mathbf{u}^{[\ell]}(\mathbf{p})
\widehat{n}_{1}(\mathbf{r},\mathbf{p}) \, , \label{Eq. 8}
\end{equation}
%
%Eq. (9)
\begin{equation}
\widehat{\mathbf{I}}_{h}^{[\ell ]}(\mathbf{r}) = \int d^{3}p
\frac{p^{2}}{2m} \frac{\mathbf{p}}{m}
\widehat{n}_{1}(\mathbf{r},\mathbf{p}) \, , \label{Eq. 9}
\end{equation}
%
%Eq. (10)
\begin{equation}
\widehat{I}_{h}^{[\ell ]}(\mathbf{r}) = \int d^{3}p \frac{p^{2}}{2m}
\, \mathbf{u}^{[\ell]}(\mathbf{p})
\widehat{n}_{1}(\mathbf{r},\mathbf{p}) \, , \label{Eq. 10}
\end{equation}
where
%
%Eq. (11)
\begin{equation}
\mathbf{u}^{[\ell]}(\mathbf{p}) = \left[ \frac{\mathbf{p}}{m}:...(
\ell - \mathrm{times} )...: \frac{\mathbf{p}}{m}\right]  \label{Eq.
11}
\end{equation}
is a $\ell$-order tensor (the inner product of $\ell$-times the
velocity $\mathbf{p}/m$) with $\ell =2,3,\ldots$.

Hence, according to the formalism
%
%Eq. (12)
\begin{equation}
\varrho_{\varepsilon }(t) = \exp \Big\{\ln \bar{\varrho}(t,0) -
\int_{-\infty}^{t} dt^{\prime} e^{\varepsilon (t^{\prime} - t)}
\frac{d}{dt^{\prime}} \ln \bar{\varrho}(t^{\prime},t^{\prime }-t)
\Big\} \, ,  \label{Eq. 12}
\end{equation}
where $\bar{\varrho}$ is an auxiliary (also dubbed instantaneous
quasi-equilibrium) probability distribution
%
%Eq. (13)
\begin{eqnarray}
\bar{\varrho}(t_{1},t_{2}) &=& \exp \Big\{- \phi (t_{1}) - \notag \\
&& \int d^{3}r[\,F_{n}({\mathbf{r}},t) \widehat{n}({\mathbf{r})} +
F_{h}({\mathbf{r}},t)\widehat{h}({
\mathbf{r})]} +  \notag \\
&& \mathbf{F}_{n}({\mathbf{r}},t) \cdot
\widehat{\mathbf{I}}_{n}(\mathbf{r}) +
\mathbf{F}_{h}({\mathbf{r}},t) \cdot
\widehat{\mathbf{I}}_{h}(\mathbf{r})+
\notag \\
&& \sum \limits_{\ell \geqslant 2}F_{n}^{[\ell]}({\mathbf{r}},t)
\odot \widehat{I}_{n}^{[\ell ]}(\mathbf{r}) + \notag \\
&& \sum\limits_{\ell \geqslant 2}F_{h}^{[\ell ]}({\mathbf{r}},t)
\odot \widehat{I}_{h}^{[\ell]}(\mathbf{r})] \Big\} \, .  \label{Eq.
13}
\end{eqnarray}
where are associated (or conjugated) with the basic hydrodynamic
variables, the nonequilibrium thermodynamic variables
%
%Eq. (14)
\begin{eqnarray}
\Bigl\{ F_{n}({\mathbf{r}},t), F_{h}({\mathbf{r}},t),
\mathbf{F}_{n}({\mathbf{r}},t), \mathbf{F}_{h}({\mathbf{r}},t),
\{F_{n}^{[\ell ]}({\mathbf{r}},t) \}, \notag \\ \{
F_{h}^{[\ell]}({\mathbf{r}},t) \} \Bigr\} \, , \label{Eq. 14}
\end{eqnarray}
and $\odot$ stands for a fully contracted product of tensors. We
recall that $\varepsilon$ is a real infinitesimal number that goes
to $+0$ after the calculation of average values have been performed.
The contribution involved accounts for historicity and
irreversibility in the process (see Appendix A). An initial time in
the remote past ($t_{0}\rightarrow -\infty$) has been used, implying
an adiabatic switch-on of relaxation processes.

We can now formulate the basic equations of NESEF-based MHT, that
is, the equations of evolution for the average values of the basic
microvariables of Eqs. (4) to (10); we denote the basic
macrovariables as
%
%Eq. (15)
\begin{equation}
\left\{ n(\mathbf{r},t), \, h(\mathbf{r},t),
\mathbf{I}_{n}(\mathbf{r},t), \, \mathbf{I}_{h}(\mathbf{r},t),
\{I_{n}^{[\ell]}(\mathbf{r},t)\},\{I_{h}^{[\ell]}(\mathbf{r},t) \}
\right\}  , \label{Eq. 15}
\end{equation}
$\ell = 2,3, \ldots$, meaning that
%
%Eq. (16)
\begin{equation}
n(\mathbf{r},t) = \mathrm{Tr} \{ \widehat{n}(\mathbf{r})
\varrho_{\varepsilon}(t) \} \, ,  \label{Eq. 16}
\end{equation}
%
%Eq. (17)
\begin{equation}
h(\mathbf{r},t) = \mathrm{Tr} \{ \widehat{h}(\mathbf{r})
\varrho_{\varepsilon}(t) \} \, ,  \label{Eq. 17}
\end{equation}
and so on for all the others. In this classical mechanics approach,
$\mathrm{Tr}$ stands for integration in phase space.

Hence, we have that
%
%Eq. (18)
\begin{equation}
\frac{\partial }{\partial t}I_{n}^{[\ell ]}(\mathbf{r},t) = \int
d^{3}p \, u^{[\ell]}(\mathbf{p}) \frac{\partial}{\partial t}
f_{1}(\mathbf{r},\mathbf{p};t) \, ,  \label{Eq. 18}
\end{equation}
which we call the hydrodynamic family $n$, and
%
%Eq. (19)
\begin{equation}
\frac{\partial }{\partial t}I_{h}^{[\ell]}(\mathbf{r},t) = \int
d^{3}p \, \frac{p^{2}}{2m} u^{[\ell]}(\mathbf{p})
\frac{\partial}{\partial t} f_{1}(\mathbf{r},\mathbf{p};t) \, ,
\label{Eq. 19}
\end{equation}
which we call the hydrodynamic family $h$.

Here, $\ell =0$ for the densities, $\ell =1$ for the first fluxes,
$\ell =2,3,\ldots$ for the higher-order fluxes, and
$f_{1}(\mathbf{r},\mathbf{p};t)$ is the single-particle distribution
function
%
%Eq. (20)
\begin{equation}
f_{1}(\mathbf{r},\mathbf{p};t) = \mathrm{Tr} \{
\widehat{n}_{1}(\mathbf{r},\mathbf{p}) \varrho_{\varepsilon} \left(
t\right) \} \, ,  \label{Eq. 20}
\end{equation}
which can be generically called the Boltzmann distribution function.

The distribution $f_{1}(\mathbf{r},\mathbf{p};t)$ satisfies the
single-particle kinetic equation [33] (also called the generalized
Boltzmann equation) and therefore the set of Eqs. (18), that is, the
equations of MHT are the multiple moment equations in the solution
of the generalized Boltzmann equation (initiated with the 14-moment
approach by Grad [37]). (We noticed that in theoretical statistics,
the method of moments for the solution of evolution equations of
probability distributions seems to have been originally proposed by
the renowned Russian mathematician Tchekycheff in the 19th century).
We call attention to the fact that there is a type of redundancy in
the pair of the basic equations, in the sense that
%
%Eq. (21)
\begin{equation}
I_{h}^{[\ell]}(\mathbf{r},t) = \frac{1}{2} m \mathrm{Tr}_{12}
I_{n}^{[\ell + 2]}(\mathbf{r},t) \, ,  \label{Eq. 21}
\end{equation}
where $\mathrm{Tr}_{12}$ stands for the contraction of the first two
indexes.

Inspection of Eqs. (18) and (19) clearly tells us that the evolution
equations for the hydrodynamic variables are exclusively dependent
on the evolution equation for the single-particle distribution
function $f_{1}(\mathbf{r},\mathbf{p};t)$, which is the mechanical
equation of motion for the single-particle dynamical operator
$\widehat{n}_{1}$ averaged over the nonequilibrium ensemble, namely,
%
%Eq. (22)
\begin{eqnarray}
\frac{\partial}{\partial t} f_{1}(\mathbf{r},\mathbf{p};t) &=&
\frac{\partial}{\partial t} \mathrm{Tr} \left\{
\{\widehat{n}_{1}(\mathbf{r},\mathbf{p}),\widehat{H} \}
\varrho_{\varepsilon }(t) \right\} \notag \\
&=& \mathrm{Tr} \{ [i\mathcal{L}
\widehat{n}_{1}(\mathbf{r},\mathbf{p})] \varrho_{\varepsilon}(t) \}
\, , \label{Eq. 22}
\end{eqnarray}
where $\mathcal{L}$ is the Liouville operator of the system.
However, solving these equations is a quite difficult task of almost
unmanageable proportions. Hence, this situation calls for the
derivation of an appropriate kinetic theory of practical use. This
can be done in a way that was initiated by several authors
[9,12,13,27,28] and systematized and extended by Lauck et al. [29].
In summary, Eq. (22) acquires the form of a generalization of Mori's
equation of motion with a highly nonlinear character, namely,
%
%Eq. (23)
\begin{eqnarray}
\frac{\partial }{\partial t} f_{1}(\mathbf{r},\mathbf{p};t) &=&
J_{1}^{(0)}(\mathbf{r},\mathbf{p};t) +
J_{1}^{(1)}(\mathbf{r},\mathbf{p};t) + \notag \\
&& + \sum \limits_{n \geqslant 2}
\Omega_{1}^{(n)}(\mathbf{r},\mathbf{p};t) \, , \label{Eq. 23}
\end{eqnarray}
where on the right hand side, the first contribution $J_{1}^{(0)}$
is in Mori's nomenclature a precession term, the second
$J_{1}^{(1)}$ is a term involving the action of the external and
internal forces, and the last contribution is a general collision
integral consisting of a series in increasing powers ($n\geqslant
2$) of the interaction strengths. These partial collision integrals
are fully described in Refs. [12,13,29], and here we simply note
that they involve pair collisions ($n=2$), triple collisions
($n=3$), and so on, with each one including memory and vertex
renormalization effects.

The first two terms on the right-hand side of Eq. (23) are given by
%
%Eq. (24)
\begin{equation}
J_{1}^{(0)}(\mathbf{r},\mathbf{p};t) = \mathrm{Tr} \left\{
\{\widehat{n}_{1}(\mathbf{r},\mathbf{p}),\widehat{H}_{0} \}
\bar{\varrho}(t,0) \varrho_{R}\right\} \, ,  \label{Eq. 24}
\end{equation}
%
%Eq. (25)
\begin{equation}
J_{1}^{(1)}(\mathbf{r},\mathbf{p};t) = \mathrm{Tr} \left\{
\{\widehat{n}_{1}(\mathbf{r},\mathbf{p}),\widehat{H}^{\prime} \}
\bar{\varrho}(t,0) \varrho_{R} \right\} \, ,  \label{Eq. 25}
\end{equation}
where $\{\ldots, \ldots \}$ stands for Poisson brackets, and
$\varrho _{R} $ is the statistical distribution of the external
reservoir at temperature $T_{0}$. In the lowest-order approximation,
which is generally valid, keeping the contribution to second order
in the interaction strengths, (``binary collisions" without memory)
[12,13,29,30,33], the last contribution in Eq. (23) reduces to:
%
%Eq. (26)
\begin{eqnarray}
J_{1}^{(2)}(\mathbf{r},\mathbf{p};t) &=& \int \limits_{- \infty}^{t}
dt^{\prime} e^{\varepsilon (t^{\prime} - t)} \times \notag \\
&& \mathrm{Tr} \Bigl\{ \{ \widehat{H}^{\prime}(t^{\prime
}-t)_{0},\{\widehat{H}^{\prime
},\widehat{n}_{1}(\mathbf{r},\mathbf{p})\}\} \times \notag \\
&& \bar{\varrho}(t,0) \varrho_{R} \Bigr\} +  \notag \\
&& \int \limits_{-\infty}^{t} dt^{\prime} e^{\varepsilon
(t^{\prime}-t)} \int d^{3} r^{\prime} d^{3} p^{\prime} \times \notag \\
&&  \mathrm{Tr} \left\{ \{\widehat{H}^{\prime}(t^{\prime}-t)_{0},
\widehat{n}_{1}(\mathbf{r}^{\prime},\mathbf{p}^{\prime}) \}
\bar{\varrho}(t,0) \varrho_{R} \right\} \times  \notag \\
&& \frac{\delta J_{1}^{(1)}(\mathbf{r},\mathbf{p};t)}{\delta
f_{1}(\mathbf{r}^{\prime},\mathbf{p}^{\prime };t)} \, ,  \label{Eq.
26}
\end{eqnarray}
where $\bar{\varrho}(t,0)$ is given in Eq. (13), and the nought
subindex indicates time evolution in the interaction representation.
The two contributions on the right-hand side of Eq. (26) consist of
the so-called irreducible contribution and the vertex
renormalization (effects of the forces present in $J_{1}^{(1)}$
acting at the time of the ``collision") [44]; $\delta$ stands for
functional differentiation [45].

The NESEF single-particle kinetic equation was presented in Ref.
[34], which included the interaction with a thermal bath (a multiple
Brownian particle system) and interparticle interactions given in
the so-called weak-coupling limit. We briefly describe the equation
in Appendix B, including a complete treatment of the two-particle
interaction, i.e., going beyond the weak coupling limit.

In the absence of a thermal bath, that is, considering the single
fluid described at the beginning of this Section, we have that the
kinetic equation for $f_{1}(\mathbf{r},\mathbf{p};t)$ is given by
%
%Eq. (27)
\begin{equation}
\frac{\partial}{\partial t} f_{1}(\mathbf{r},\mathbf{p};t) =
J_{1}^{(0)}(\mathbf{r},\mathbf{p};t) +
J_{1}^{(1)}(\mathbf{r},\mathbf{p};t) +
J_{1}^{(2)}(\mathbf{r},\mathbf{p};t) \, ,  \label{Eq. 27}
\end{equation}
where the first terms on the right-hand side are
%
%Eq. (28)
\begin{eqnarray}
J_{1}^{(0)}(\mathbf{r},\mathbf{p};t) &=& \mathrm{Tr} \left\{
\{\widehat{n}_{1}(\mathbf{r},\mathbf{p}),\widehat{H}_{0}\}
\bar{\varrho}(t,0) \varrho_{R}\right\}  \notag \\
&=& - \frac{\mathbf{p}}{m} \cdot \nabla
f_{1}(\mathbf{r},\mathbf{p};t) \, , \label{Eq. 28}
\end{eqnarray}
which is, as already noticed, the precession term in Moris's
terminology, and
%
%Eq. (29)
\begin{eqnarray}
J_{1}^{(1)}(\mathbf{r},\mathbf{p};t) &=& \mathrm{Tr} \left\{
\{\widehat{n}_{1}(\mathbf{r},\mathbf{p}),\widehat{H}^{\prime}\}
\bar{\varrho}(t,0) \varrho_{R} \right\}  \notag \\
&=& [\nabla U_{ex}(\mathbf{r},t) + \nabla U(\mathbf{r},t)] \cdot
\nabla_{\mathbf{p}} f_{1}(\mathbf{r},\mathbf{p};t) \notag \, , \\
\label{Eq. 29}
\end{eqnarray}
is the contribution containing the action of the external and
internal forces, where
%
%Eq. (30)
\begin{equation}
U(\mathbf{r},t) = \int d^{3}r^{\prime} d^{3}p^{\prime }
V(|\mathbf{r} - \mathbf{r}^{\prime}|)
f_{1}(\mathbf{r}^{\prime},\mathbf{p}^{\prime};t) \label{Eq. 30}
\end{equation}
plays the role of a mean-field potential of the interaction between
particles or a Vlasov-like potential, and
%
%Eq. (31)
%
\begin{equation}
J_{1}^{(2)}(\mathbf{r},\mathbf{p};t) =
J_{11}^{(2)}(\mathbf{r},\mathbf{p};t) +
J_{12}^{(2)}(\mathbf{r},\mathbf{p};t) +
J_{13}^{(2)}(\mathbf{r},\mathbf{p};t) \, ,  \label{Eq. 31}
\end{equation}
is the collision integral resulting from the interaction between
particles, shown in Appendix B. The first contribution on the
right-hand side is, we recall, the weak-coupling contribution.

Consequently, the hydrodynamic equations, Eq. (18) and (19), take
the form
%
%Eq. (32)
\begin{eqnarray}
\frac{\partial}{\partial t} I_{n}^{[\ell]}(\mathbf{r},t) &=& \int
d^{3}p\,u^{[\ell]}(\mathbf{p}) [J_{1}^{(0)}(\mathbf{r},\mathbf{p};t)
+ J_{1}^{(1)}(\mathbf{r},\mathbf{p};t) + \notag \\
&& + J_{1}^{(2)}(\mathbf{r},\mathbf{p};t)] \, ,  \label{Eq. 32}
\end{eqnarray}
%
%Eq. (33)
\begin{eqnarray}
\frac{\partial}{\partial t}I_{h}^{[\ell]}(\mathbf{r},t) &=& \int
d^{3} p\frac{p^{2}}{2m}
u^{[\ell]}(\mathbf{p})[J_{1}^{(0)}(\mathbf{r},\mathbf{p};t) + \notag \\
&& + J_{1}^{(1)}(\mathbf{r},\mathbf{p};t) +
J_{1}^{(2)}(\mathbf{r},\mathbf{p};t)] \, ,  \label{Eq. 33}
\end{eqnarray}
recalling that $\ell =0$ stands for the densities, $\ell =1$ stands
for the first fluxes, and $\ell \geqslant 2$ stands for the
higher-order fluxes.

In Section IV, we address in complete detail the case of contraction
in an MHT description of order 2, considering the $n$-family.

%%%%%%%%%%%%%%%%%%%%%%%%%%%%%%%%%%%%%%%%%%%%%%%%%%%%%%%%%%%%%%%%%%%%%%%%%%
%%%%%%%%%%%%%%%%%%%%%%%%%%%%%%%%%%%%%%%%%%%%%%%%%%%%%%%%%%%%%%%%%%%%%%%%%%
%%%%%%%%%%%%%%%%%Section 3%%%%%%%%%%%%%%%%%%%%%%%%%%%%%%%%%%%%%%%%%%%%%%%%
%%%%%%%%%%%%%%%%%%%%%%%%%%%%%%%%%%%%%%%%%%%%%%%%%%%%%%%%%%%%%%%%%%%%%%%%%%

\section{Contraction of Description}

On the issue of the contraction of the description, we noticed that
a truncation criterion can be derived, which rests on the
characteristics of the hydrodynamic motion that develops under the
given experimental procedure.

Since inclusion of higher- and higher-order fluxes implies
describing motion with increasing Knudsen numbers for each
hydrodynamic mode (that is governed by smaller and smaller
wavelengths -- larger and larger wavenumbers -- accompanied by
higher and higher frequencies), in a qualitative manner, we can say,
as a general \textquotedblleft rule of thumb," that the criterion
indicates that \emph{an increasingly restricted contraction can be
used when the prevalent wavelengths in the motion are larger}.
Therefore, in simpler words, when the motion becomes smoother in
space and time, the dimension of the space of basic macrovariables
used for the description of the nonequilibrium thermodynamic state
of the system can be further reduced.

As shown elsewhere [42], a general contraction criterion can be
conjectured, namely, a truncation of order $r$ (keeping the
densities and their fluxes to order $r$) can be introduced, once we
show that in the spectrum of wavelengths, the motion predominates
over the "frontier" motion, $\lambda_{(r,r+1)}^{2}=v^{2}\theta
_{r}\theta_{r+1}$, where $v$ is of the order of the thermal
velocity, and $\theta_{r}$ and $\theta_{r+1}$ are the corresponding
Maxwell times associated with the $r$- and $r+1$-fluxes [38,46,47].
We recall that the Maxwell time was originally introduced by Maxwell
in his fundamental article of 1867 [48] on the dynamical theory of
gases in what was related to viscoelasticity. A family of Maxwell
times is present in MHT associated with the dampening of particle
and energy densities and their fluxes of all orders [38].

%%%%%%%%%%%%%%%%%%%%%%%%%%%%%%%%%%%%%%%%%%%%%%%%%%%%%%%%%%%%%%%%%%%%%
%%%%%%%%%%%%%%%%%   Section 4  %%%%%%%%%%%%%%%%%%%%%%%%%%%%%%%%%%%%%%
%%%%%%%%%%%%%%%%%%%%%%%%%%%%%%%%%%%%%%%%%%%%%%%%%%%%%%%%%%%%%%%%%%%%%
\section{Mesoscopic Hydro-Thermodynamics of Order 2}

We consider a contraction of the description of order 2, MHT[2], in
the evolution of the family $n$, that is, we introduce the set of
hydrodynamics variables
%
%Eq. (34)
\begin{equation}
\left\{ n(\mathbf{r},t), \, \mathbf{I}_{n}(\mathbf{r},t), \,
I_{n}^{[2]}(\mathbf{r},t) \right\} \, ,  \label{Eq. 34}
\end{equation}
a thirteen-moments approach (or Grad's fourteen moments once the
energy is incorporated in $I^{[2]}$, which is related to the
pressure tensor; see Eq. (21)). We call attention to the fact that
the approach here has a purely mechanical basis, in a laboratory
reference frame, while the Grad's approach is a hybrid approach
including hydrodynamic variables in a barycentric frame of reference
in the definitions.

Recalling that the second flux contains the density of energy,
namely, (cf. Eq. (21))
%
%Eq. (35)
\begin{equation}
h(\mathbf{r},t) = m \mathrm{Tr} \left\{ I_{h}^{[2]}(\mathbf{r},t)
\right\} \, , \label{Eq. 35}
\end{equation}
it is convenient to improve the physical discussions to redefine the
set of basic hydrodynamic variables as
%
%Eq. (36)
\begin{equation}
\left\{ h(\mathbf{r},t), n(\mathbf{r},t),
\mathbf{I}_{n}(\mathbf{r},t) \, \mathring{I}_{n}^{[2]}(\mathbf{r},t)
\right\} \, ,  \label{Eq. 36}
\end{equation}
where we have introduced the traceless part of the second flux, that
is, $\mathring{I}_{n}^{[2]}(\mathbf{r},t)$.

The conjugated nonequilibrium thermodynamic variables, see Appendix
A, are designed by
%
%Eq. (37)
\begin{equation}
\left\{ F_{h}({\mathbf{r}},t), F_{n}({\mathbf{r}},t),
\mathbf{F}_{n}({\mathbf{r}},t),
\mathring{F}_{n}^{[2]}({\mathbf{r}},t) \right\} \, ,  \label{Eq. 37}
\end{equation}
which are redefined as
%
%Eq. (38)
\begin{equation}
F_{h}({\mathbf{r}},t) \equiv \beta ({\mathbf{r}},t) =
\frac{1}{k_{B}T^{\ast}({\mathbf{r}},t)} \, ,  \label{Eq. 38}
\end{equation}
%
%Eq. (39)
\begin{equation}
F_{n}({\mathbf{r}},t) \equiv - \beta ({\mathbf{r}},t)
\mu^{\ast}({\mathbf{r}},t) \, ,  \label{Eq. 39}
\end{equation}
%
%Eq. (40)
\begin{equation}
\mathbf{F}_{n}({\mathbf{r}},t) \equiv - \beta ({\mathbf{r}},t)
\mathbf{v}({\mathbf{r}},t) \, ,  \label{Eq. 40}
\end{equation}
introducing a nonequilibrium temperature (usually called
quasi-temperature), $T^{\ast}({\mathbf{r}},t)$, where
$\mathbf{v}({\mathbf{r}},t)$ is the field of the barycentric
velocity, and $\mu^{\ast}({\mathbf{r}},t)$ is a nonequilibrium
chemical potential [49,50]. Moreover, we write
%
%Eq. (41)
\begin{equation}
n(\mathbf{r},t) = n_{0} + \triangle n(\mathbf{r},t) \, ,  \label{Eq.
41}
\end{equation}
%
%Eq. (42)
\begin{equation}
T^{\ast}(\mathbf{r},t) = T_{0} + \triangle T^{\ast}(\mathbf{r},t) \,
, \label{Eq. 42}
\end{equation}
where $n_{0}$ and $T_{0}$ are values in equilibrium, and we admit
that $\triangle n(\mathbf{r},t) \ll n_{0}$ and $\triangle
T^{\ast}(\mathbf{r},t)\ll T_{0}$.

Using the Heims-Jaynes perturbation expansion for averages [51] to
\emph{first order} (see Appendix C), we obtain the nonequilibrium
equations of state, which consist of a linear relation between the
nonequilibrium thermodynamic variables in Eq. (35) and the basic
variables in Eq. (34), given by
%
%Eq. (43)
\begin{equation}
h(\mathbf{r},t) = \frac{3}{2} n(\mathbf{r},t) T^{\ast}(\mathbf{r},t)
\, , \label{Eq. 43}
\end{equation}
%
%Eq. (44)
\begin{equation}
\mathbf{I}_{n}(\mathbf{r},t) = n(\mathbf{r},t)
\mathbf{v}(\mathbf{r},t) \simeq n_{0}\mathbf{v}(\mathbf{r},t) \, ,
\label{Eq. 44}
\end{equation}
%
%Eq. (45)
\begin{equation}
\,\mathring{I}_{n}^{[2]}(\mathbf{r},t) \simeq
\frac{2(k_{B})^{2}}{m^{2}}
n_{0}\mathring{F}_{n}^{[2]}({\mathbf{r}},t) \, ,  \label{Eq. 45}
\end{equation}

We recall that the evolution equations are
%
%Eq. (46)
\begin{equation}
\frac{\partial }{\partial t}n(\mathbf{r},t) =
J_{n}^{(0)}(\mathbf{r},t) + J_{n}^{(1)}(\mathbf{r},t) +
J_{n}^{(2)}(\mathbf{r},t) \, , \label{Eq. 46}
\end{equation}
%
%Eq. (47)
\begin{equation}
\frac{\partial }{\partial t}\mathbf{I}_{n}(\mathbf{r},t) =
J_{\mathbf{I}_{n}}^{(0)}(\mathbf{r},t) +
J_{\mathbf{I}_{n}}^{(1)}(\mathbf{r},t) +
J_{\mathbf{I}_{n}}^{(2)}(\mathbf{r},t) \, , \label{Eq. 47}
\end{equation}
%
%Eq. (48)
\begin{equation}
\frac{\partial }{\partial t}I_{n}^{[2]}(\mathbf{r},t) =
J_{I_{n}}^{[2](0)}(\mathbf{r},t) + J_{I_{n}}^{[2](1)}(\mathbf{r},t)
+ J_{I_{n}}^{[2](2)}(\mathbf{r},t) \, . \label{Eq. 48}
\end{equation}

Lengthy calculations lead to cumbersome expressions that we omit
here. We introduce the following approximations:

1. In the expressions of the kinetic coefficients, we take
$n(\mathbf{r},t) \simeq n_{0}$ and $T^{\ast}(\mathbf{r},t) \simeq
T_{0}$ (cf. Eqs. (39) and (40)).

2. Spatial correlations are neglected; that is, a local
approximation is introduced, and the set of coupled nonlinear
integro-differential equations is greatly simplified. Moreover, of
the three contributions to the collision integral registering the
two-particle interaction, only one contributes to the weak-coupling
interactions; the others become null because of symmetry effects. We
recall that weak coupling involves collisions with a low-angle
transfer of momentum.

The evolution equation for the density is
%
%Eq. (49)
\begin{equation}
\frac{\partial}{\partial t} n(\mathbf{r},t) + \nabla \cdot
\mathbf{I}_{n}(\mathbf{r},t) = 0 \, ,  \label{Eq. 49}
\end{equation}
which is the conservation equation for the number of molecules. For
the first flux, we obtain that
%
%Eq. (50)
\begin{eqnarray}
\frac{\partial }{\partial t}\mathbf{I}_{n}(\mathbf{r},t) & \simeq &
- \nabla \cdot I_{n}^{[2]}(\mathbf{r},t) - \frac{n_{0}}{m} \nabla
\overline{V}(\mathbf{r},t) + \notag \\
&& - a_{10} \nabla n(\mathbf{r},t) + a_{12} \nabla \cdot
I_{n}^{[2]}(\mathbf{r},t) + \notag \\
&& - \theta_{1}^{-1} \mathbf{I}_{n}(\mathbf{r},t) \, ,   \label{Eq.
50}
\end{eqnarray}
where
%
%Eq. (51)
\begin{equation}
\overline{V}(\mathbf{r},t) = \int d^{3}r^{\prime
}V(\mathbf{r}-\mathbf{r}^{\prime})n(\mathbf{r}^{\prime },t)
\label{Eq. 51}
\end{equation}
is the mean field (Vlasov-Landau-like field) of the two-particle
interactions,
%
%Eq. (52)
\begin{equation}
a_{10} = \frac{7}{5} n_{0} \frac{\mathcal{A}}{mk_{B}T_{0}} \, ,
\label{Eq. 52}
\end{equation}
%
%Eq. (53)
\begin{equation}
a_{12} = \frac{2}{5} n_{0} \frac{\mathcal{A}}{(k_{B}T_{0})^{2}} \, ,
\label{Eq. 53}
\end{equation}
%
%Eq. (54)
\begin{equation}
\mathcal{A} = \frac{1}{4 \pi^{2}} \int dq d\theta |\Phi(q)|^{2}
q^{3} \cos^{2} \theta \sin \theta \, ,  \label{Eq. 54}
\end{equation}
where $\Phi (q)$ is the Fourier transform of the two-particle
potential, and $\theta_{1}$ is the Maxwell time associated with the
first flux given by
%
%Eq. (55)
\begin{equation}
\theta_{1}^{-1} = \mathcal{A} \frac{n_{0}}{k_{B}T_{0}} \sqrt{\frac{2
\pi}{mk_{B}T_{0}}} \, .  \label{Eq. 55}
\end{equation}

In Eq. (48), the first term on the right-hand side comes from
$J_{I}^{(0)}$, the second term $J_{I}^{(1)}$ contains the force
arising out of the mean field originating in the two-molecule
interaction, and the other three terms have their origin in
$J_{n}^{(2)}$. The first two terms are contributions from a
self-energy correction involving the gradient of the previous term,
that is, $\nabla n(\mathbf{r},t)$, and the divergence of the next
term, that is, $\nabla \cdot I_{n}^{[2]}(\mathbf{r},t)$ (quite
analogously to those present in Ref. [5], here given in an
expression at the microscopic-statistical level), and the last term
can be referred to as a generalized Maxwellian relaxation that
introduces a generalized Maxwell time. The origin of the Maxwell
time goes back to a fundamental article by J. C. Maxwell in 1867
[48], interpreted as the time during which stresses in media are
damped [51]; each flux in MHT has an associated Maxwell time, and
the relaxation times of the hydrodynamic motion are a combination of
these multiple Maxwell times.

For the second-order flux, we recall that it contains the density of
energy (cf. Eq. (35)), and we have that
%
%Eq. (56)
\begin{eqnarray}
\frac{\partial}{\partial t} I_{n}^{[2]}(\mathbf{r},t) &=& - \nabla
\cdot I_{n}^{[3]}(\mathbf{r},t) \notag \\
&& - \frac{1}{m}\left\{
[\mathbf{I}_{n}(\mathbf{r},t) \colon \nabla \overline{V}(\mathbf{r},t)] +
transposed \right\}  \notag \\
&& + a_{20} n(\mathbf{r},t)1^{[2]} + a_{21} \nabla \cdot
\mathbf{I}_{n}(\mathbf{r},t) 1^{[2]}  \notag \\
&& - \frac{1}{4} a_{12} \left\{ \nabla \colon
\mathbf{I}_{n}(\mathbf{r},t) + transposed \right\} \notag \\
&& - \theta_{2}^{-1} I_{n}^{[2]}(\mathbf{r},t) \, , \label{Eq. 56}
\end{eqnarray}
where
%
%Eq. (57)
\begin{equation}
a_{20} = \frac{k_{B}T_{0}}{m} \theta_{2}^{-1} \, ,  \label{Eq. 57}
\end{equation}
%
%Eq. (58)
\begin{equation}
a_{21} = \frac{4}{5} \frac{\mathcal{B}}{mk_{B}T_{0}} \, , \label{Eq.
58}
\end{equation}
and $\theta_{2}$ is Maxwell time associated with the second-order
flux given by
%
%Eq. (59)
\begin{equation}
\theta_{2}^{-1} = \frac{3}{5} \theta_{1}^{-1} \, .  \label{Eq. 59}
\end{equation}

Moreover, taking into account that (cf. Eq. (21)) the density of
energy is
%
%Eq. (60)
\begin{equation}
h(\mathbf{r},t) = \frac{1}{2} m \mathrm{Tr} \left\{
I_{n}^{[2]}(\mathbf{r},t) \right\} \, ,  \label{Eq. 60}
\end{equation}
%
%Eq. (61)
\begin{equation}
\mathbf{I}_{h}(\mathbf{r},t) = \frac{1}{2} m \mathrm{Tr}_{12}
\left\{ I_{n}^{[3]}(\mathbf{r},t) \right\} \, ,  \label{Eq. 61}
\end{equation}
resorting to the use of Eqs. (46), (48), (55) and that
%
%Eq. (62)
\begin{eqnarray}
\nabla \left\{ \nabla \cdot I_{n}^{[3]}(\mathbf{r},t)\right\}
&\simeq& \frac{k_{B}T_{0}}{m} \bigl\{ \nabla^{2}
\mathbf{I}_{n}(\mathbf{r},t)
+ \notag \\
&& 2 \nabla (\nabla \cdot \mathbf{I}_{n}(\mathbf{r},t)) \bigr\}
I_{n}^{[2]}(\mathbf{r},t) \, ,  \label{Eq. 62}
\end{eqnarray}
after using the Heims-Jaynes formalism in the linear approximation,
we obtain the evolution equation for the density energy given by
%
%Eq. (63)
\begin{eqnarray}
\frac{\partial}{\partial t}h(\mathbf{r},t) + \nabla
I_{h}(\mathbf{r},t) &=& - \nabla \overline{V}(\mathbf{r},t) \cdot
\mathbf{I}_{n}(\mathbf{r},t) + \notag \\
&& + \frac{3}{2} k_{B}T_{0} \theta_{2}^{-1} n(\mathbf{r},t) +  \notag \\
&& + n_{0} \frac{\mathcal{B}}{k_{B}T_{0}} \nabla \cdot
\mathbf{I}_{n}(\mathbf{r},t) + \notag \\
&& - \theta_{2}^{-1} h(\mathbf{r},t) \, , \label{Eq. 63}
\end{eqnarray}
and according to Eq. (59), the momentum relaxation time is smaller
than the energy relaxation time.

However, after some mathematical handling of Eqs. (46), (48), (55)
and using Eq. (61), we obtain for the density that
%
%Eq. (64)
\begin{equation*}
\frac{\partial ^{3}n(\mathbf{r},t)}{\partial t^{3}} +
\frac{1}{\theta _{ef}} \frac{\partial ^{2}n(\mathbf{r},t)}{\partial
t^{2}} + \frac{1}{(\theta _{(2)})^{2}}\frac{\partial
n(\mathbf{r},t)}{\partial t} +
\end{equation*}
\begin{equation*}
-\frac{C_{GNL}^{2}}{\theta _{2}}\nabla ^{2}n(\mathbf{r},t) =
\end{equation*}
\begin{eqnarray}
&=& \frac{3n_{0}\mathcal{B}}{5mk_{B}T_{0}} \nabla ^{2}[\nabla \cdot
\mathbf{I}_{n}(\mathbf{r},t)] + \notag \\
&& - \left( 1 - \frac{2n_{0}\mathcal{B}}{5(k_{B}T_{0})^{2}} \right)
\frac{k_{B}T_{0}}{m} \times \notag \\
&& \nabla \cdot \lbrack \nabla^{2}
\mathbf{I}_{n}(\mathbf{r},t) + 2 \nabla (\nabla \cdot \mathbf{I}_{n}(\mathbf{r},t))]  \notag \\
&& - \frac{n_{0}\mathcal{B}}{5mk_{B}T_{0}} \nabla \cdot \{\nabla
\cdot \lbrack \nabla \mathbf{I}_{n}(\mathbf{r},t) + (\nabla
\mathbf{I}_{n}(\mathbf{r},t)^{tr})] \} \notag \, , \\  \label{Eq.
64}
\end{eqnarray}
where we discarded terms in $\mathcal{B}^{2}$,
%
%Eq. (65)
\begin{equation}
C_{GNL}^{2} = \frac{k_{B}T_{0}}{m} +
\frac{n_{0}\mathcal{B}}{mk_{B}T_{0}} = C_{GL}^{2} +
\frac{n_{0}\mathcal{B}}{mk_{B}T_{0}} \, , \label{Eq. 65}
\end{equation}
%
%Eq. (66)
\begin{equation}
C_{GL}^{2} = \frac{k_{B}T_{0}}{m} + \frac{n_{0}
\mathcal{B}}{mk_{B}T_{0}} = C_{GL}^{2} + \frac{n_{0}
\mathcal{B}}{mk_{B}T_{0}} \, , \label{Eq. 66}
\end{equation}
%
%Eq. (67)
\begin{equation}
\frac{1}{\theta_{ef}} = \frac{1}{\theta_{1}} + \frac{1}{\theta_{2}}
\qquad \mathrm{and} \qquad \theta_{(2)} = \sqrt{\theta_{1}
\theta_{2}} \, , \label{Eq. 67}
\end{equation}

This equation has a third-order differentiation in time: neglecting
the third derivative leads to a generalized Maxwell-Cattaneo
equation, and if the second derivative is neglected, we have a
generalized Fick diffusion equation. This implies hydrodynamic
motion that becomes smoother in time.

%%%%%%%%%%%%%%%%%%%%%%%%%%%%%%%%%%%%%%%%%%%%%%%%%%%%%%%%%%%%%%%%%%%%%%%%%%
%%%%%%%%%%%%%%%    Section 6    %%%%%%%%%%%%%%%%%%%%%%%%%%%%%%%%%%%%%%%%%%
%%%%%%%%%%%%%%%%%%%%%%%%%%%%%%%%%%%%%%%%%%%%%%%%%%%%%%%%%%%%%%%%%%%%%%%%%%

\section{Concluding Remarks}

After an introduction with a brief discussion of some aspects of
hydrodynamics, we presented so-called mesoscopic
hydro-thermodynamics, or higher-order generalized hydrodynamics,
built in the framework of a mechanical-statistical formalism (the
non-equilibrium statistical ensemble formalism).

MHT (HOGH) allows, in principle, to describe hydrodynamic motion
without restrictions in the values of the wavelengths and
frequencies involved. The theoretical description involves including
the densities of particles and energy together with their fluxes of
all orders, which are related by an enormous set of coupled
nonlinear integro-differential equations describing the motion. The
solution is not practically feasible, and one needs to introduce a
contraction of the description, that is, one needs to retain a
finite small number of fluxes, up to a certain order $n$, considered
to be appropriate for describing the motion under consideration. The
order of the contracted description is determined by a criterion
(see Section III), which heavily depends on the values of the
Maxwell-like times that are associated with the different fluxes.
These times are clearly characterized in the equations of motion.

It may be noted that this is a completely analytical theory, and
competing theories are computational modeling theories derived from
the nonequilibrium molecular dynamics formalism; they seem to lead
to comparable numerical results.

\appendix{}

%%%%%%%%%%%%%%%%%%%%%%%%%%%%%%%%%%%%%%%%%%%%%%%%%%%%%%%%%%%%%%%%%%%%%
%%%%%%%%%%%%%  Appendix A  %%%%%%%%%%%%%%%%%%%%%%%%%%%%%%%%%%%%%%%%%%
%%%%%%%%%%%%%%%%%%%%%%%%%%%%%%%%%%%%%%%%%%%%%%%%%%%%%%%%%%%%%%%%%%%%%

\section{The Nonequilibrium Statistical Operator}

The construction of nonequilibrium statistical ensembles, that is, a
nonequilibrium statistical ensemble formalism, NESEF for short,
which essentially consists of a derivation of a nonequilibrium
statistical operator (probability distribution in the classical
case), has been attempted along several lines. In a brief summarized
way, we describe the construction of the NESEF within a heuristic
approach. First, it should be noticed that for systems away from
equilibrium, several important points need to be carefully taken
into account in each case under consideration:

\begin{enumerate}
\item \emph{The choice of the basic variables} (a wholly different
choice from the case of equilibrium when it suffices to take a set
of variables that are constants of motion), which is based on an
analysis of what sort of macroscopic measurements and processes
are actually possible; moreover, one should focus attention not
only on what can be observed but also on the character and expectation
concerning the equations of evolution for these variables [12,13,25].

\item \emph{The question of irreversibility} (or Eddington's arrow of time);

\item \emph{Historicity needs be introduced}, that is, the idea that the
past dynamics of the system (or historicity effects) must be incorporated
along the time interval going from an initial description of the macrostate
of the sample in a given experiment, say at $t_{0}$, to the time $t$ when a
measurement is performed.
\end{enumerate}

Concerning the choice of the basic variables, in contrast to the
case of equilibrium, immediately after an open system of $N$
particles, in contact with external sources and reservoirs, has been
driven out of equilibrium, it becomes necessary to describe its
state in terms of all the observables and, eventually, introduce
direct and cross-correlation. However, as time elapses, Bogoliubov's
principle of correlation weakening allows us to introduce increasing
contractions of the descriptions. Let us say that we can introduce a
description based on the observables $\{\hat{P}_{j}\}$, $j = 1, 2
..., n$, on which the nonequilibrium statistical operator depends.

On the question of irreversibility, in the absence of a proper way
to introduce such an effect, one needs to resort to the
\emph{interventionist's approach}, which is based on the ineluctable
process of randomization leading to the asymmetric evolution of the
macrostate.

The ``intervention" consists of introducing into the Liouville
equation of the statistical operator, of an otherwise isolated
system, a particular source accounting for Krylov's
``\emph{jolting}" \emph{effect} [52], in the form (written for the
logarithm of the statistical operator)
%
%Eq. (A.1)
\begin{equation}
\frac{\partial}{\partial t} \ln \mathcal{R}_{\varepsilon }(t) +
\frac{1}{ i \hbar }[\ln \mathcal{R}_{\varepsilon }(t),\hat{H}] = -
\varepsilon \lbrack \ln \mathcal{R}_{\varepsilon}(t) - \ln
\overline{\mathcal{R}}(t,0)], \label{Eq. A1}
\end{equation}
where $\varepsilon$ (a kind of reciprocal of a relaxation time) is
taken to go to $+0$ after the calculations of average values have
been performed. Such a mathematically inhomogeneous term, in the
otherwise normal Liouville equation, implies a continuous tendency
of relaxation of the statistical operators towards a
\emph{referential distribution}, $\overline{\mathcal{R}}$, which
represents an instantaneous quasi-equilibrium condition.

We can see that Eq. (A.1) consists of a regular Liouville equation
but with an infinitesimal source, which provides Bogoliubov's
symmetry breaking of time reversal and is responsible for
disregarding the advanced solutions. This is described by a Poisson
distribution, and the result at time $t$ is obtained by averaging
over all $t^{\prime}$ in the interval $(t_{0},t)$, and the solution
of Eq. (A.1) is
%
%Eq. (A.2)
\begin{equation}
\mathcal{R}_{\varepsilon }(t) = \exp \left\{ - \hat{S}(t,0) + \int
\limits_{t_{0}}^{t}dt^{\prime} e^{\varepsilon (t^{\prime} - t)}
\frac{d}{dt^{\prime}} \hat{S}(t^{\prime},t^{\prime} - t) \right\} \,
, \label{Eq. A2}
\end{equation}
where
%
%Eq. (A.3)
\begin{equation}
\hat{S}(t,0) = - \ln \overline{\mathcal{R}}(t,0) \, ,  \label{Eq.
A3}
\end{equation}
%
%Eq. (A.4)
\begin{eqnarray}
\hat{S}(t^{\prime},t^{\prime} - t) &=& \exp \left\{ -\frac{1}{i
\hbar}(t^{\prime} - t) \hat{H} \right\} \hat{S}(t^{\prime},0) \times  \notag \\
&& \exp \left\{ \frac{1}{i \hbar}(t^{\prime} - t) \hat{H} \right\}
\, , \label{Eq. A4}
\end{eqnarray}
and the initial-time condition at time $t_{0}$, when the application
of the formalism begins, is
%
%Eq. (A.5)
\begin{equation}
\mathcal{R}_{\varepsilon }(t_{0}) = \overline{\mathcal{R}}(t_{0},0)
\, . \label{Eq. A5}
\end{equation}

In $\overline{\mathcal{R}}$ and $\hat{S}$, the first time variable
in the argument refers to the evolution of the nonequilibrium
thermodynamic variables, and the second time variable refers to the
time evolution of the dynamical variables, both of which have an
effect on the operator. The statistical operator can be written in
the form
%
%Eq. (A.6)
\begin{equation}
\mathcal{R}_{\varepsilon }(t) = \overline{\mathcal{R}}(t,0) +
\mathcal{R}_{\varepsilon}^{\prime}(t) \, .  \label{Eq. A6}
\end{equation}
involving the auxiliary probability distribution
$\bar{\varrho}(t,0)$, with $\varrho_{\varepsilon}^{\prime}(t)$,
which contains the historicity and irreversibility effects.
Moreover, in most cases, we can consider a system that consist of
the system of interest (on which we are performing an experiment) in
contact with ideal reservoirs. Thus, we can write
%
%Eq. (A.7)
\begin{equation}
\overline{\mathcal{R}}(t,0) = \bar{\varrho}(t,0) \times \varrho _{R}
\, . \label{Eq. A7}
\end{equation}
and
%
%Eq. (A.8)
\begin{equation}
\mathcal{R}_{\varepsilon }(t) = \varrho_{\varepsilon}(t) \times
\varrho_{R} \, ,  \label{Eq. A8}
\end{equation}
where $\varrho_{\varepsilon}(t)$ is the statistical operator of the
nonequilibrium system, $\bar{\varrho}$ is the auxiliary operator,
and $\varrho_{R}$ is the stationary operator of the ideal
reservoirs, with $\varrho_{\varepsilon}(t)$ given by
%
%Eq. (A.9)
\begin{equation}
\varrho_{\varepsilon}(t) = \exp \left\{ - \hat{S}(t,0) + \int
\limits_{- \infty}^{t} dt^{\prime} e^{\varepsilon (t^{\prime} - t)}
\frac{d}{dt^{\prime}} \hat{S}(t^{\prime},t^{\prime} - t) \right\} \,
, \label{Eq. A9}
\end{equation}
with the initial value $\bar{\varrho}(t_{0},0)$ ($t_{0}\rightarrow -
\infty$), and where
%
%Eq. (A.10)
\begin{equation}
\hat{S}(t,0) = - \ln \bar{\varrho}(t,0) \, .  \label{Eq. A10}
\end{equation}

Finally, the auxiliary statistical operator $\bar{\varrho}(t,0)$
must be provided. This operator defines an instantaneous
distribution at time $t$, which describes a \textquotedblleft
frozen" equilibrium defining the macroscopic state of the system at
the given time, and for that reason the operator is sometimes dubbed
as the \emph{quasi-equilibrium statistical operator}. On this basis
(or, alternatively, via the variational procedure [12,13,14]) and
considering the description of the nonequilibrium state of the
system in terms of the basic set of dynamical variables
$\hat{P}_{j}$, the reference or instantaneous quasi-equilibrium
statistical operator is taken as a canonical-like operator given by
%
%Eq. (A.11)
\begin{equation}
\bar{\varrho}(t,0) = \exp \Big\{ - \phi (t) - \sum \limits_{j}^{n}
F_{j}(t) \hat{P}_{j} \Big\} \, ,  \label{Eq. A11}
\end{equation}
with $\phi (t)$ ensuring the normalization of $\bar{\varrho}$ and
playing the role of a logarithm of a partition function, say, $\phi
(t) = \ln \bar{Z}(t)$. Moreover, in Eq. (A.11), $F_{j}$ are the
nonequilibrium thermodynamic variables associated with each kind of
basic dynamical variable $\hat{P}_{j}$. The nonequilibrium
thermodynamic space of states consists of the basic variables
$\{Q_{j}(t)\}$, which consist of averages of $\{\hat{P}_{j}\}$ over
the nonequilibrium ensemble, namely,
%
%Eq. (A.12)
\begin{equation}
Q_{j}(t) = \mathrm{Tr} \{\hat{P}_{j} \varrho_{\varepsilon}(t)\} \, ,
\label{Eq. A12}
\end{equation}
which are then functionals of $\{F_{j}(t)\}$, and the equations of
state are
%
%Eq. (A.13)
\begin{equation}
Q_{j}(t) = - \frac{\delta \phi (t)}{\delta F_{j}(t)} = -
\frac{\delta \ln \bar{Z}(t)}{\delta F_{j}(t)} \, ,  \label{Eq. A13}
\end{equation}
where $\delta$ stands for a functional derivative [45].

Moreover,
%
%Eq. (A.14)
\begin{equation}
\bar{S}(t) = \mathrm{Tr} \{ \hat{\bar{S}}(t,0) \bar{\varrho}(t,0) \}
= - \mathrm{Tr} \{ \bar{\varrho}(t,0) \ln \bar{\varrho}(t,0)\} \, ,
\label{Eq. A14}
\end{equation}
is the so-called informational entropy characteristic of the
distribution $\bar{\varrho}$, a functional of the basic variables
$\{Q_{j}(t)\}$, and the alternative form of the equations of state
is given by
%
%Eq. (A.15)
\begin{equation}
-\frac{\delta \bar{S}(t)}{\delta Q_{j}(t)} = F_{j}(t) \, .
\label{Eq. A15}
\end{equation}

In the contracted description used for the construction of HOGH of
order 2, that is, in terms of the variables of Eq. (36), we have
that the auxiliary statistical operator, better written in
reciprocal space, is
%
%Eq. (A.16)
%
\begin{eqnarray}
\bar{\varrho}(t,0) &=& \exp \Big\{ - \phi(t) \notag \\
&& - \sum \limits_{\mathbf{Q}} [F_{h}({\mathbf{Q}},t)
\widehat{h}({\mathbf{Q})
+ F_{n}({\mathbf{Q}},t)\widehat{n}({\mathbf{Q})}} +  \notag \\
&& + \mathbf{F}_{n}(\mathbf{Q},t) \cdot
\widehat{\mathbf{I}}_{n}(\mathbf{Q}){+}
\mathring{F}_{n}^{[2]}(\mathbf{Q},t) \otimes
\widehat{\mathring{I}}_{n}^{[2]}(\mathbf{Q}) \notag \, . \\
\label{Eq. A16}
\end{eqnarray}

Quantities with a triangular hat are the mechanical operators
corresponding to the quantities in Eq. (3). In Eq. (A.16), the
conjugated nonequilibrium thermodynamical variables, $F$'s listed in
Eq. (37), are introduced.

%%%%%%%%%%%%%%%%%%%%%%%%%%%%%%%%%%%%%%%%%%%%%%%%%%%%%%%%%%%%%%%%%%%%%%%%%%%%%%%%%%
%%%%%%%%%%%%%Appendix B%%%%%%%%%%%%%%%%%%%%%%%%%%%%%%%%%%%%%%%%%%%%%%%%%%%%%%%%%%%
%%%%%%%%%%%%%%%%%%%%%%%%%%%%%%%%%%%%%%%%%%%%%%%%%%%%%%%%%%%%%%%%%%%%%%%%%%%%%%%%%%
\section{Single-Particle Kinetic Equation}

The NESEF-based single-particle kinetic equation was reported in
Ref. [34], which contained an interaction with a surrounding thermal
bath and assumed that in a dilute solution the interparticle
interaction could be treated in the weak-coupling approximation. We
reproduce below this result but incorporate the collision integral
associated with the two-particle interaction in full. We have
%
%Eq. (B1)
\begin{equation}
\frac{\partial }{\partial t}f_{1}(\mathbf{r},\mathbf{p};t) =
J_{1}^{(0)}(\mathbf{r},\mathbf{p};t) +
J_{1}^{(1)}(\mathbf{r},\mathbf{p};t) +
J_{1}^{(2)}(\mathbf{r},\mathbf{p};t) \, ,  \label{Eq. B1}
\end{equation}
where the first terms on the right-hand side are
%
%Eq. (B2)
\begin{eqnarray}
J_{1}^{(0)}(\mathbf{r},\mathbf{p};t) &=& \mathrm{Tr} \left\{
\{\widehat{n}_{1}(\mathbf{r},\mathbf{p}), \widehat{H}_{0} \}
\bar{\varrho}(t,0) \varrho_{R} \right\}  \notag \\
&=& - \frac{\mathbf{p}}{m} \cdot \nabla
f_{1}(\mathbf{r},\mathbf{p};t) \, , \label{Eq. B2}
\end{eqnarray}
which are, as already noticed, the precession term in Moris's
terminology, and
%
%Eq. (B3)
\begin{eqnarray}
J_{1}^{(1)}(\mathbf{r},\mathbf{p};t) &=& \mathrm{Tr} \left\{
\{\widehat{n}_{1}(\mathbf{r},\mathbf{p}), \widehat{H}^{\prime} \}
\bar{\varrho}(t,0) \varrho_{R} \right\}  \notag \\
&=& [\nabla U_{ex}(\mathbf{r},t) + \nabla U(\mathbf{r},t)] \cdot
\nabla_{\mathbf{p}} f_{1}(\mathbf{r},\mathbf{p};t) \notag \, , \\
\label{Eq. B3}
\end{eqnarray}
is the contribution containing the action of the external and
internal forces, where
%
%Eq. (B4)
\begin{equation}
U(\mathbf{r},t) = \int d^{3}r^{\prime} d^{3}p^{\prime} V(|\mathbf{r}
- \mathbf{r}^{\prime}|)
f_{1}(\mathbf{r}^{\prime},\mathbf{p}^{\prime};t) \label{Eq. B4}
\end{equation}
plays the role of a mean-field potential of the interaction between
particles or a Vlasov-like potential. Observe that the dot stands
for the scalar product of vectors, and we write $\odot$ for the full
contraction of tensors.

Taking all the interactions in Eq. (26), denoted by $H^{\prime}$,
that is, pair interactions with applied external sources, lengthy
but straightforward calculations give the final expression of the
collision integral $J_{1}^{(2)}$. Taking this final expression, Eqs.
(B2) and (B3)\ into Eq. (B1) leads to the kinetic equation:
%
%Eq. (B5)
\begin{equation*}
\frac{\partial}{\partial t} f_{1}(\mathbf{r},\mathbf{p};t) +
\frac{\mathbf{P}(\mathbf{r},\mathbf{p};t)}{m} \cdot \nabla
f_{1}(\mathbf{r},\mathbf{p};t) +
\end{equation*}
%
%Eq. (B5)
\begin{equation*}
+ \mathbf{F}(\mathbf{r},\mathbf{p};t) \cdot \nabla_{\mathbf{p}}
f_{1}(\mathbf{r},\mathbf{p};t) - B(\mathbf{p})
f_{1}(\mathbf{r},\mathbf{p};t) +
\end{equation*}
\begin{equation*}
-A_{2}^{[2]}(\mathbf{p}) \odot \lbrack \nabla_{\mathbf{p}} \nabla]
f_{1}(\mathbf{r},\mathbf{p};t) - B_{2}^{[2]}(\mathbf{p}) \odot
\lbrack \nabla_{\mathbf{p}} \nabla_{\mathbf{p}}]
f_{1}(\mathbf{r},\mathbf{p};t) =
\end{equation*}
\begin{equation}
= J_{S}^{(2)}(\mathbf{r},\mathbf{p};t) \, ,  \label{Eq. B5}
\end{equation}
where
%
%Eq. (B6)
\begin{equation}
\mathbf{P}(\mathbf{p};t) = \mathbf{p} -m \mathbf{A}_{1}(\mathbf{p})
\, , \label{Eq. B6}
\end{equation}
plays the role of a generalized momentum,
%
%Eq. (B7)
%
\begin{equation}
\mathbf{F}(\mathbf{r},\mathbf{p};t) = - \nabla U_{ex}(\mathbf{r},t)
- \mathbf{B}_{1}(\mathbf{p}) - \mathbf{F}_{NL}(\mathbf{r},t) -
\nabla U(\mathbf{r},t) \, , \label{Eq. B7}
\end{equation}
is a generalized force, in which
%
%Eq. (B8)
%
\begin{equation}
\mathbf{F}_{NL}(\mathbf{r};t) = \int d^{3}r^{\prime} \int
d^{3}p^{\prime} \mathbf{G}_{NL}(\mathbf{r}^{\prime} -
\mathbf{r},\mathbf{p}^{\prime})
f_{1}(\mathbf{r}^{\prime},\mathbf{p}^{\prime};t) \, ,  \label{Eq.
B8}
\end{equation}
and the coefficients are as follows:
%
%Eq. (B9)
%
\begin{equation}
\mathbf{A}_{1}(\mathbf{p}) = - \frac{n_{B}}{\mathcal{V}} \frac{M
\beta_{0}}{m} \sum_{\mathbf{Q}}\mathbf{Q} \frac{\left\vert \psi
(Q)\right\vert ^{2}}{Q^{2}} \mathbf{Q} \cdot \nabla _{\mathbf{p}}
F(\mathbf{p,Q}) \, ,  \label{Eq. B9}
\end{equation}
where
%
%Eq. (B10)
\begin{equation}
F(\mathbf{p,Q}) = 1 + \sum
\limits_{n=1}^{\infty}\frac{(-1)^{n}}{(2n-1)!!} \left( \frac{M
\beta_{0}}{Q^{2}m^{2}} \right)^{n} \left( \mathbf{Q} \cdot
\mathbf{p} \right)^{2n} \, ,  \label{Eq. B10}
\end{equation}
%

%
%Eq. (B11)
\begin{eqnarray}
\mathbf{B}_{1}(\mathbf{p}) &=& \frac{n_{B}}{\mathcal{V}} \frac{\pi
\sqrt{(M \beta_{0})^{3}}}{m\sqrt{2\pi}} \times \notag \\
&& \sum_{\mathbf{Q}}\mathbf{Q} \frac{\left\vert \psi (Q)\right\vert
^{2}}{Q} \left( \frac{1}{M} - \frac{1}{m}\right) \left( \mathbf{Q}
\cdot \mathbf{p} \right) e^{- \frac{M \beta _{0} \left( \mathbf{Q}
\cdot \mathbf{p} \right)^{2}}{2Q^{2}m^{2}}} \notag \, , \\
\label{Eq. B11}
\end{eqnarray}
%

%
%Eq. (B12)
\begin{equation*}
\mathbf{G}_{NL}(\mathbf{r}^{\prime} -
\mathbf{r},\mathbf{p}^{\prime}) = \frac{n_{R}
\beta_{0}}{\mathcal{V}} \sum \limits_{\mathbf{Q}}\mathbf{Q}
\left\vert \psi (Q) \right\vert^{2} \Big\{ i
F(\mathbf{Q},\mathbf{p}^{\prime}) +
\end{equation*}
\begin{equation}
+\left( \frac{M\beta _{0}}{2\pi} \right)^{1/2} \frac{\pi}{m}
\frac{\mathbf{Q} \cdot \mathbf{p}^{\prime}}{Q} e^{ - \alpha \left(
\frac{\mathbf{Q}}{Q} \cdot \mathbf{p}^{\prime }\right) ^{2}}\Big\}
e^{i \mathbf{Q \cdot (r}^{\prime} - \mathbf{r)}}\,, \label{Eq. B12}
\end{equation}
%

%
%Eq. (B13)
\begin{equation}
A_{2}^{[2]}(\mathbf{p}) = \frac{n_{B}}{\mathcal{V}} \frac{M
\beta_{0}}{m} \sum_{\mathbf{Q}} \frac{\left\vert \psi (Q)
\right\vert ^{2}}{Q^{2}} F(\mathbf{p,Q}) [\mathbf{QQ}] \, ,
\label{Eq. B13}
\end{equation}
%

%
%Eq. (B14)
\begin{equation}
B_{2}^{[2]}(\mathbf{p}) = \frac{n_{B}}{\mathcal{V}} \sqrt{ \frac{M
\beta_{0}}{2 \pi}} \pi \sum_{\mathbf{Q}} \frac{ \left\vert \psi (Q)
\right\vert ^{2}}{Q} e^{ - \frac{M\beta _{0}\left( \mathbf{Q} \cdot
\mathbf{p}\right)^{2}}{2Q^{2}m^{2}}} [\mathbf{QQ}] \, ,  \label{Eq.
B14}
\end{equation}
%

%
%Eq. (B15)
\begin{eqnarray}
B(\mathbf{p}) &=& \frac{n_{B}}{\mathcal{V}} \frac{\pi \sqrt{(M
\beta_{0})^{3}}}{m M \sqrt{2\pi}} \times \notag \\
&& \sum_{\mathbf{Q}}\frac{\left\vert \psi (Q) \right\vert ^{2}}{Q}
\left( Q^{2} - \frac{M \beta_{0}}{m^{2}} \left( \mathbf{Q} \cdot
\mathbf{p} \right)^{2} \right) e^{ - \frac{M \beta_{0} \left(
\mathbf{Q} \cdot \mathbf{p} \right)^{2}}{2Q^{2}m^{2}}} \notag \, . \\
\label{Eq. B15}
\end{eqnarray}

In Eqs. (B.13) and (B.14) $[\mathbf{QQ}]$ stands for a tensor
product of vectors, i.e., a rank-2 tensor. Moreover, in these
equations, $\psi (Q)$ is the Fourier transform of the potential
energy $w(|(\mathbf{r}_{j}-\mathbf{R}_{\mu })|$, $\mathcal{V}$ is
the volume and $n_{B}=N_{B}/\mathcal{V}$.

%%%%%%%%%%%%%%%%%%%%%%%%%%%%%%%%%%%%%%%%%%%%%%%%%%%%%%%%%%%%%%%%%%%%%%%%%%%%%%%%
%%%%%%%%%%%%%%%%%%%   Appendix C   %%%%%%%%%%%%%%%%%%%%%%%%%%%%%%%%%%%%%%%%%%%%%
%%%%%%%%%%%%%%%%%%%%%%%%%%%%%%%%%%%%%%%%%%%%%%%%%%%%%%%%%%%%%%%%%%%%%%%%%%%%%%%%

\section{Summary of Heims-Jaynes Procedure}

Given a statistical operator of the form
%
%Eq. (C1)
\begin{equation}
\varrho = \frac{1}{Z} e^{\widehat{A} + \widehat{B}} , \label{Eq. C1}
\end{equation}
where
%
%Eq. (C2)
\begin{equation}
Z = \mathrm{Tr} \big\{ e^{\widehat{A} + \widehat{B}} \big\} \, ,
\label{Eq. C2}
\end{equation}
ensures its normalization and introducing
%
%Eq. (C3)
\begin{equation}
\varrho_{0} = \frac{e^{\widehat{A}}}{\mathrm{Tr} \big\{
e^{\widehat{A}} \big\} } \, , \label{Eq. C3}
\end{equation}
according to Heims-Jaynes, given an operator $\widehat{\Theta}$, it
follows that
%
%Eq. (C4)
\begin{equation}
\mathrm{Tr}\big\{\widehat{\Theta }\varrho \big\} = \langle
\widehat{\Theta} \rangle_{0} + \sum \limits_{n=1}^{\infty} \langle
\widehat{Q}_{n}(\widehat{\Theta} - \langle \widehat{\Theta} \rangle
_{0})\rangle \, , \label{Eq. C4}
\end{equation}
where
%
%Eq. (C5)
\begin{equation}
\langle \widehat{\Theta} \rangle_{0} = \mathrm{Tr} \big\{
\widehat{\Theta} \varrho_{0} \big\} \, , \label{Eq. C5}
\end{equation}
with
%
%Eq. (C6)
\begin{equation}
\widehat{Q}_{n} = \widehat{S}_{n} - \sum \limits_{k=1}^{n-1} \langle
\widehat{Q}_{n} \rangle_{0}\widehat{S}_{n-k} \, , \label{Eq. C7}
\end{equation}
for $n\geq 2$, and $\widehat{Q}_{0}=\widehat{1}$ and
$\widehat{Q}_{1} = \widehat{S}_{1}$,
%
%Eq. (C7)
\begin{equation}
\widehat{S}_{n} = \frac{B^{n}}{n!} \, , \quad \widehat{S}_{0} =
\widehat{1} \, . \label{Eq. C7}
\end{equation}

Equation (C4) consists of the average value of $\widehat{\Theta}$
with $\varrho_{0}$ (that is, only depending on $A$) plus a
contribution in the form of a series expansion in powers of $B$. In
a first-order approximation, we have that
%
%Eq. (C8)
\begin{equation}
\mathrm{Tr} \big\{ \widehat{\Theta} \varrho \big\} \simeq \langle
\widehat{\Theta} \rangle_{0} + \mathrm{Tr} \big\{
\widehat{B}(\widehat{\Theta} - \langle \widehat{\Theta} \rangle_{0})
\varrho_{0}\big\} \, . \label{Eq. C8}
\end{equation}

In Section III, we have used
%
%Eq. (C9)
\begin{equation}
\overline{\varrho}(t,0) = \frac{1}{\overline{Z}(t)} e^{\widehat{A} +
\widehat{B}} \, , \label{Eq. C9}
\end{equation}
where
%
%Eq. (C10)
\begin{equation}
\widehat{A}(t) = - \sum_{\mathbf{Q}} F_{h}(\mathbf{Q},t)
\widehat{h}(\mathbf{Q}) \, , \label{Eq. C10}
\end{equation}
%
%Eq. (C11)
\begin{eqnarray}
\widehat{B}(t) &=& - \sum_{\mathbf{Q}} [F_{n}(\mathbf{Q},t)
\widehat{n}(\mathbf{Q}) + \mathbf{F}_{n}(\mathbf{Q},t) \cdot
\widehat{\mathbf{I}}_{n}(\mathbf{Q}) + \notag \\
&& + \mathring{F}_{n}^{[2]}(\mathbf{Q},t) \otimes
\widehat{I}_{n}^{[2]}(\mathbf{Q})] \, , \label{Eq. C11}
\end{eqnarray}
and
%
%Eq. (C12)
\begin{equation}
\overline{Z}(t)=e^{\phi (t)} \, , \label{Eq. C12}
\end{equation}
with the calculations performed in the first-order (linear)
Heims-Jaynes expansion.

%%%%%%%%%%%%%%%%%%%%%%%%%%%%%%%%%%%%%%%%%%%%%%%%%%%%%%%%%%%%%%%%%%%%%%%%%%%%%%%%
%%%%%%%%%%%%%%%%%%%%%%Bibliography%%%%%%%%%%%%%%%%%%%%%%%%%%%%%%%%%%%%%%%%%%%%%%%%
%%%%%%%%%%%%%%%%%%%%%%%%%%%%%%%%%%%%%%%%%%%%%%%%%%%%%%%%%%%%%%%%%%%%%%%%%%%%%%%%

\end{document}